\documentclass[12pt]{iopart}
 \usepackage{epsfig}

\def\gsim{\,\raise0.3ex\hbox{$>$\kern-0.75em\raise-1.1ex\hbox{$\sim$}}\,}
\def\lsim{\,\raise0.3ex\hbox{$<$\kern-0.75em\raise-1.1ex\hbox{$\sim$}}\,}
\begin{document}

\title{Hadron multiplicities, $p_T$ spectra
 and net-baryon number in central Pb+Pb collisions at the LHC}
\vspace{-2mm}
 
\author{KJ Eskola,$\hspace{-1mm}$  H Honkanen,$\hspace{-1mm}^a$ H Niemi, PV Ruuskanen, SS 
R\"as\"anen$^b$}

\address{Department of Physics, University of Jyv\"askyl\"a, Finland\\ 
Helsinki Institute of Physics, University of Helsinki, Finland\\
$^a$Department of Physics, University of Virginia, Charlottesville, VA, USA\\
$^b$Jyv\"askyl\"a University of Applied Sciences, Jyv\"askyl\"a, Finland
}

We summarize here our recent LHC predictions \cite{EHNRR}, obtained in the framework of perturbative 
QCD (pQCD)+saturation+hydrodynamics (EKRT model for brief) \cite{EKRT}. This model has successfully 
predicted \cite{EKRT,ERRT} the charged particle multiplicities in central Au+Au collisions at 
different $\sqrt{s}$, and it also describes the low-$p_T$ spectra of pions and kaons at RHIC quite 
well \cite{EHNRR,ENRR}. 

Primary parton production in the EKRT model is computed from collinearly factorized pQCD cross 
sections \cite{EKL89} by extending the calculation towards smaller $p_T$ until the abundant gluon 
production vertices overlap and gluon fusions \cite{GLR} saturate the number of produced partons 
(gluons). The saturation scale is determined as $p_0=p_{\rm sat}$ from a saturation condition 
\cite{EKRT}
$
N_{AA}(p_0, \sqrt s)\cdot \pi/p_0^2  = c \cdot \pi R_A^2,
$
where $N_{AA}(p_0, \sqrt s)$ is the average number of partons produced at $|y|\le0.5$ and 
$p_T\ge p_0$. With a constant $c=1$ the framework is closed. For central Pb+Pb collisions at the LHC
$p_{\rm sat}\approx 2$~GeV. We obtain the initial conditions for the cylindrically symmetric boost 
invariant 2+1-D hydrodynamical description by converting the computed transverse energy $E_T(p_{\rm 
sat})$ and net-baryon number $N_{B}(p_{\rm sat})$ into densities $\epsilon(r,\tau_0)$ and 
$n_B(r,\tau_0)$ using binary collision profiles and formation time $\tau_0=1/p_{\rm sat}$.

Assuming a fast thermalization at $\tau_0$, and zero initial transverse fluid
velocity, we proceed by solving the standard equations of ideal
hydrodynamics including the current conservation equation for net-baryon
number. In the Equation of State we assume an ideal gas of gluons and
massless quarks ($N_f=3$), the QGP, with a bag constant $B$ at
$T>T_c$, and a hadron resonance gas of all states with $m<2$ GeV at
$T<T_c$. Taking $B^{1/4}=239$ MeV leads to 1st-order transition with
$T_c=165$ MeV. Final state hadron spectra are obtained with Cooper-Frye
procedure on a decoupling surface at $T_{\rm dec}$ followed by
strong and electromagnetic 2- and 3-body decays of unstable states using
the known branching ratios. Extensive comparison \cite{EHNRR,ENRR} with RHIC data suggests
a single decoupling temperature $T_{\rm dec}=150$ MeV which is also used
to calculate the predictions for the LHC. For details, see \cite{EHNRR}.

Our predictions \cite{EHNRR} for the LHC multiplicities, transverse energies and net-baryon number at 
$y=\eta=0$ for 5 \% most central Pb+Pb collisions at $\sqrt s=5.5$~TeV are summarized in the table 
below. Note that the predicted charged particle multiplicity $dN_{\rm ch}/d\eta$ is 2570, i.e. only a 
third of the initial ALICE design value (see also \cite{ERRT}).
Whereas the multiplicity of initially produced partons and observable
hadrons are close to each other, the transverse energy is reduced by a
factor as large as 3.4 in the evolution from initial state to final
hadrons. Due to this reduction the very high initial temperature, $T_0\gsim 1$~GeV, 
possibly observable through the emission of photons, need not lead to contradiction 
between predicted and observed $E_T$.
\begin{table}[hbt]
\lineup
\begin{tabular}{@{}*{13}{l}}
\br                              
$\frac{dN}{dy}^{\rm tot}$&$\frac{dN}{d\eta}^{\rm tot}$&$\frac{dN}{dy}^{\rm ch}$& 
$\frac{dN}{d\eta}^{\rm ch}$&$\frac{dN}{dy}^{B}$&$\frac{dE}{dy}^{T}$&$\frac{dE}{d\eta}^{T}$& 
$\frac{dN}{dy}^{\pi\pm}$&$\frac{dN}{dy}^{\pi0}$
&$\frac{dN}{dy}^{K\pm}$&$\frac{dN}{dy}^{p}$&$\frac{dN}{dy}^{\bar p}$&$p/\bar p$
\cr 
\mr
{\small 4730} 					& {\small 4240} 						& {\small 2850}					
&\textbf{{\small 2570}}					&
{\small 3.11}					& {\small 4070}						& {\small 3710}
& {\small 1120} & {\small 1240} & {\small 214} & {\small 70.8} & {\small 69.6} & {\small 0.98}
\cr
\br
\end{tabular}
\vspace{-4mm}
\end{table}

Our prediction for the charged hadron $p_T$ spectrum is the lower limit 
of the red band (HYDRO, the width corresponding to $T_{\rm dec}=120\dots 150$~MeV) in the l.h.s. 
figure below \cite{EHNRR}. The corresponding $p_T$ distributions  of $\pi^+$ and $K^+$ are shown in 
the r.h.s. figure (solid lines). The pQCD reference spectra, obtained by folding the LO pQCD cross 
sections with the nuclear PDFs and  fragmentation functions (KKP) and accounting for the NLO 
contributions with a $\sqrt s$-dependent $K$-factor from \cite{EH03}, are also shown (pQCD) on the 
r.h.s.. The yellow bands (pQCD w E-loss) show the results with parton energy losses included as in 
\cite{EHSW04}. We thus predict the applicability region of hydrodynamics at the LHC to be $p_T\lsim 
4\dots 5$~GeV, i.e. a wider region than at RHIC.  
\begin{figure}[tbh]
\vspace{-5mm}
 \begin{flushleft}
   \vspace{-1.4cm}
    \epsfxsize 7.5cm \epsfbox{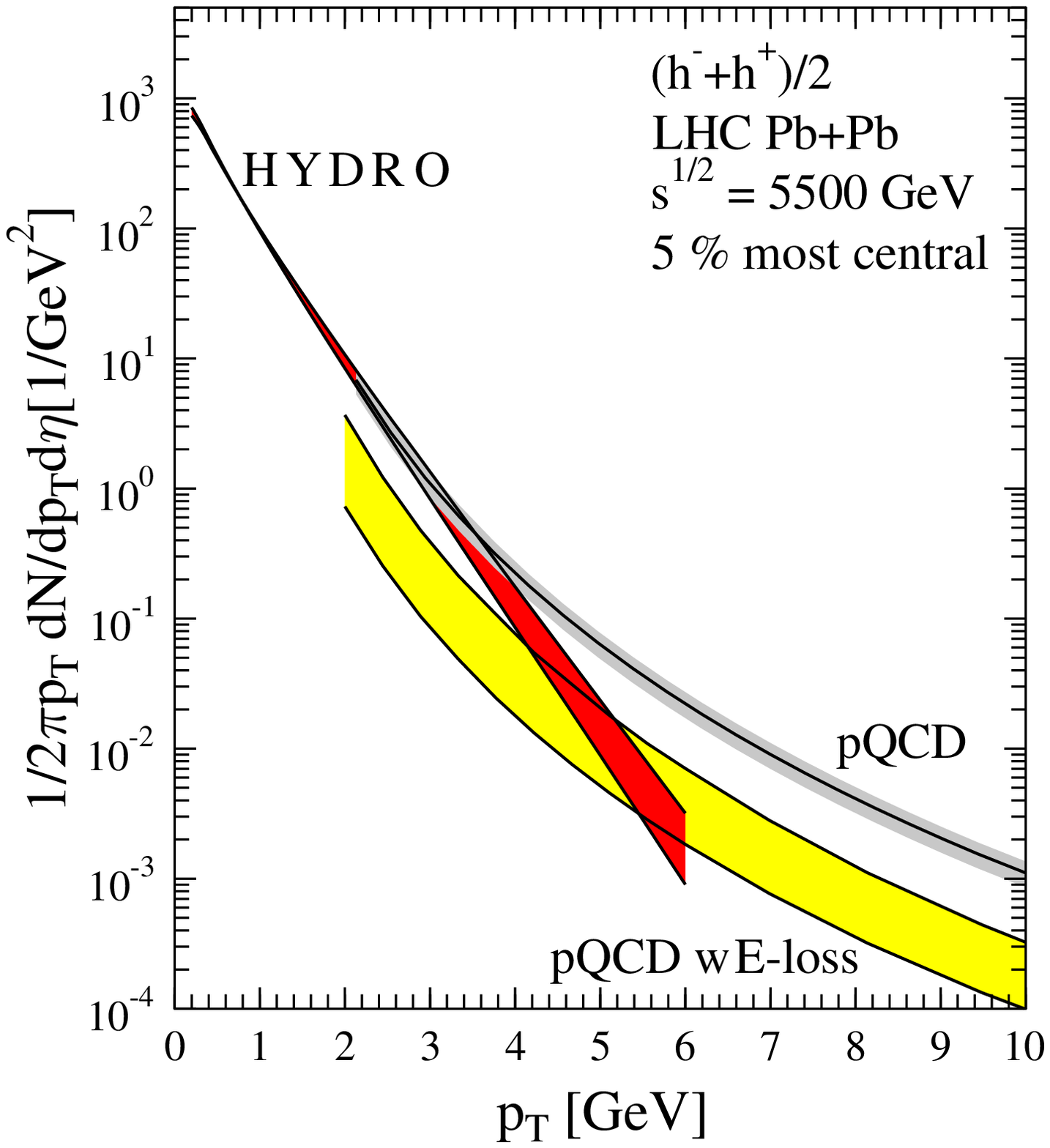} 
\end{flushleft}
\begin{flushright}
	\vspace{-8.66cm}
	\epsfxsize 7.5cm \epsfbox{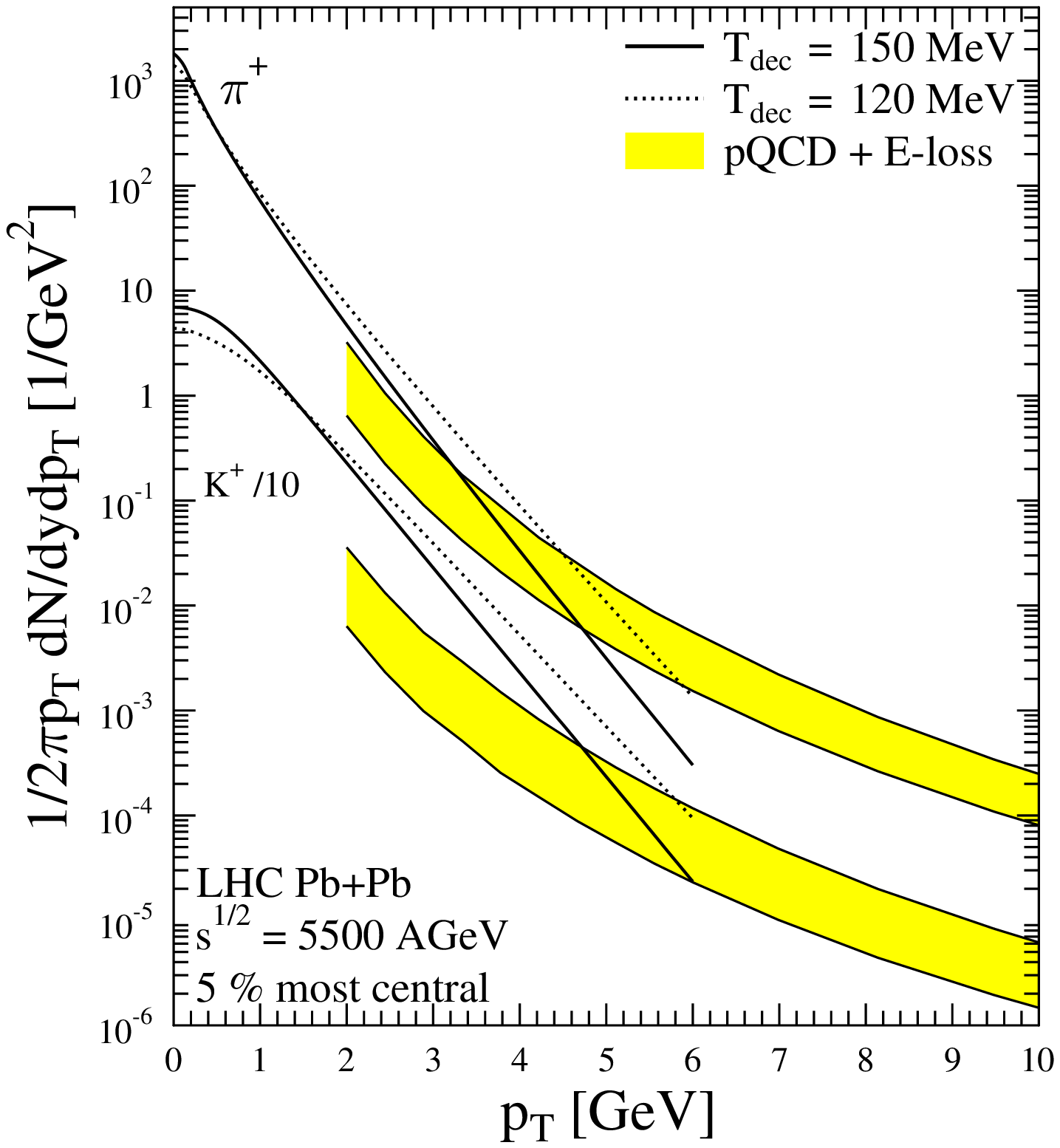}
   \vspace{-1.2cm}
  \label{fig:chargedLHC}
 \end{flushright}

\vspace{-2cm}
\end{figure}


\section*{References}
\begin{thebibliography}{10}
\vspace{-2mm}
\bibitem{EHNRR}
K.~J.~Eskola, H.~Honkanen, H.~Niemi, P.~V.~Ruuskanen and S.~S.~R\"as\"anen,
Phys.\ Rev.\ C {\bf 72} (2005) 044904 [arXiv:hep-ph/0506049]

\bibitem{EKRT}
K.~J.~Eskola, K.~Kajantie, P.~V.~Ruuskanen and K.~Tuominen,
Nucl.\ Phys.\ B {\bf 570} (2000) 379
[arXiv:hep-ph/9909456].

\bibitem{ERRT}
K.~J.~Eskola, P.~V.~Ruuskanen, S.~S.~R\"as\"anen and K.~Tuominen,
Nucl.\ Phys.\ A {\bf 696} (2001) 715 [arXiv:hep-ph/0104010].

\bibitem{ENRR}
K.~J.~Eskola, H.~Niemi, P.~V.~Ruuskanen and S.~S.~R\"as\"anen,
Phys.\ Lett.\ B {\bf 566} (2003) 187-192 [arXiv:hep-ph/0206230].

\bibitem{EKL89}
K.~J.~Eskola, K.~Kajantie and J.~Lindfors,
Nucl.\ Phys.\ B {\bf 323} (1989) 37.

\bibitem{GLR}
L.~V.~Gribov, E.~M.~Levin and M.~G.~Ryskin,
Phys.\ Rept.\ {\bf 100} (1983) 1.

\bibitem{EH03}
K.~J.~Eskola and H.~Honkanen,
Nucl.\ Phys.\ A {\bf 713} (2003) 167
[arXiv:hep-ph/0205048].

\bibitem{EHSW04}
K.~J.~Eskola, H.~Honkanen, C.~A.~Salgado and U.~A.~Wiedemann,
Nucl.\ Phys.\ A {\bf 747} (2005) 511
[arXiv:hep-ph/0406319].


\end{thebibliography}
\end{document}